# Possible correlation between cosmic ray variation and atmosphere parameters (present status and future activities at BEO Moussala and Alomar observatory)[*]

Alexander Mishev

On behalf of BEO Moussala

Institute for Nuclear Research and Nuclear Energy, Bulgarian Academy of Sciences, Basic Environmental Observatory Moussala

72 Tsarigradsko chaussee, Sofia 1284, Bulgaria

Corresponding author: Alexander Mishev
e-mail: mishev@inrne.bas.bg
Tel: ++359 2 9746310
Fax: ++3592 9753619

Abstract: *In this revue are presented several present activities and experiments at Alomar observatory and BEO Moussala and the possibility to investigate the influence of cosmic ray on climate parameters and the impact of cosmic ray to climate change. The general aim of these experiments is to study the possible correlation between cosmic ray variation and atmosphere parameters. The potential to study the possible correlations is discussed. The existing experiments and devices are presented as well these in preparation and commissioning. The scientific potential of future and in preparation experiments is discussed. The possibility to study the space weather at BEO Moussala and Alomar observatory presented. The specific devises precisely the muon telescope, neutron flux-meter, Cherenkov light telescope and muon hodoscope are presented.*

## 1. Introduction

In this section briefly is presented the cosmic ray physics basics, with the corresponding registration techniques and knowing problems.

The field of astroparticle physics is in plain extension in the last decades. This field is connected with high energy physics and gamma-ray astrophysics and crossroads both of them. Since the last several years one has been able to observe phenomena which have significant impact to our knowledge of the universe such as supernova remnants, blazars, active galactic nuclei, gamma ray bursts etc...

Evermore almost after one century of cosmic ray studies there are still many basic important unsolved problems connected with the origin and acceleration mechanisms of primary cosmic ray flux.

The primary cosmic rays extend over twelve decades of energy with the corresponding decline in the intensity. The flux goes down from $10^4$ m$^{-2}$ s$^{-1}$ at energies $\sim 10^9$ eV to $10^{-2}$ km$^{-2}$ yr$^{-1}$ at energies $\sim 10^{20}$ eV. The shape of the spectrum is with observed small deviation from the power law function across this large energy interval. In fact the change in the slope $\propto E^{-2.7}$ to $\propto E^{-3.0}$ around $1\text{-}3.10^{15}$ eV is well known as the "knee" of the spectrum (fig.1) [1]. The recent results from measurements and experiments are the discussed problems connected with mass composition and energy spectrum are presented in [2, 3]. In the region of extremely high energy [4] the problems are more or less similar.

---

[*] This work is supported under NATO grant EAP.RIG. 981843 and FP6 project BEOBAL.

The second change of the slope, actually a flattening, is observed at energies near to $10^{18}$ eV is known as "ankle" of the spectrum. The "knee" is usually associated with an energy limit of acceleration mechanisms of supernova remnants and may be related to a loss of ability for galactic magnetic field to retain the cosmic ray flux. The possible explanation is based on the super nova remnants diffuse shock acceleration mechanism [5]. Generally in this model the galactic supernovae are the only galactic candidate with sufficient energy [6]. The issue within this model is that the supernova diffuse shock acceleration mechanism can produce high energy particles up to some maximal energy, which is limited by the lifetime of the shockwave.

The "ankle" is usually associated with the onset of the dominant extra galactic cosmic ray spectrum. In this model our galaxy produces particles with energies ut to those of the "ankle". The mass composition of the primary cosmic ray flux has an aparent change as well from medium to heavy nuclei and to light nuclei in cited above regions of the energy spetra. The cosmic ray studies are complementary to gamma ray astrophysics since many gamma rays are produced in processes such as synchrotron emission as example, which involve charged cosmic ray particles.

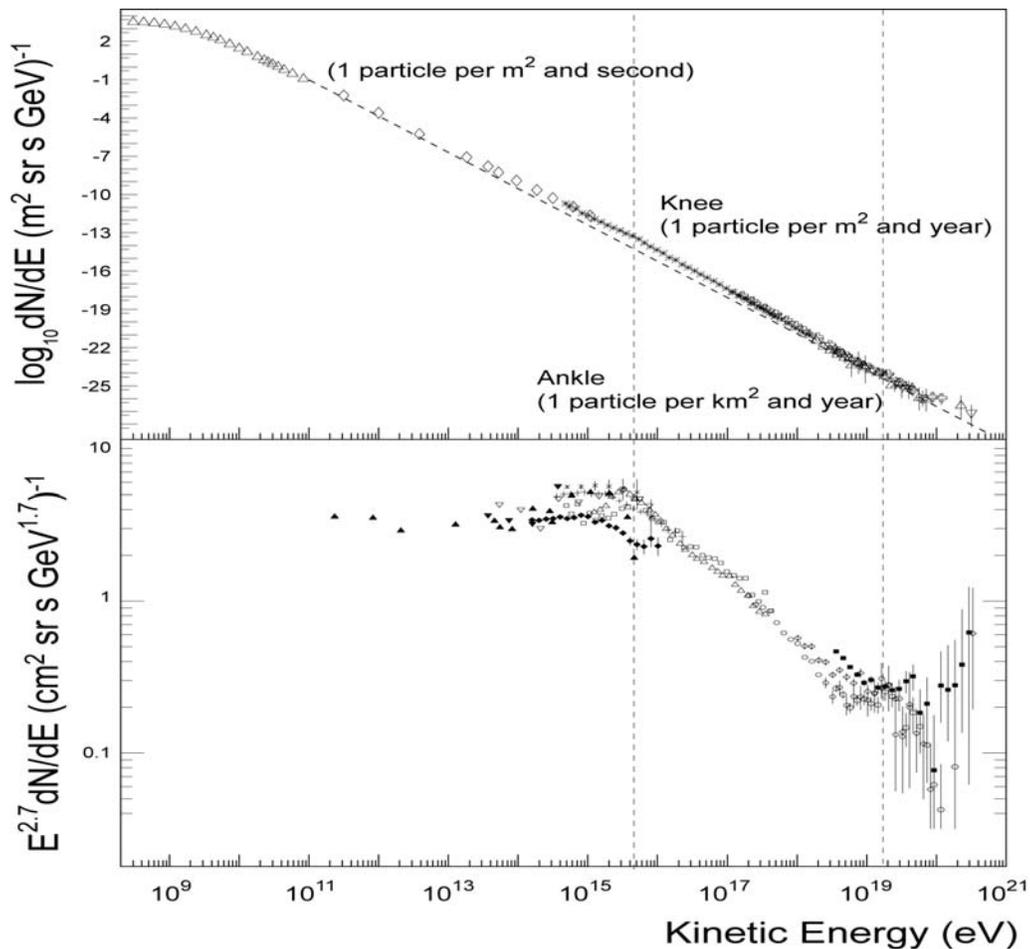

Fig.1 Primary Cosmic ray spectrum

The measurements of the individual cosmic ray spectrum (fig. 2) and the precise estimation of mass composition are very important in attempt to obtain more defined information about the sources of primary cosmic ray and to build an adequate model of cosmic ray origin [7]. In fig.2 [8] are presented the major components as a function of the energy of the primary particle i.e. the individual spectra.

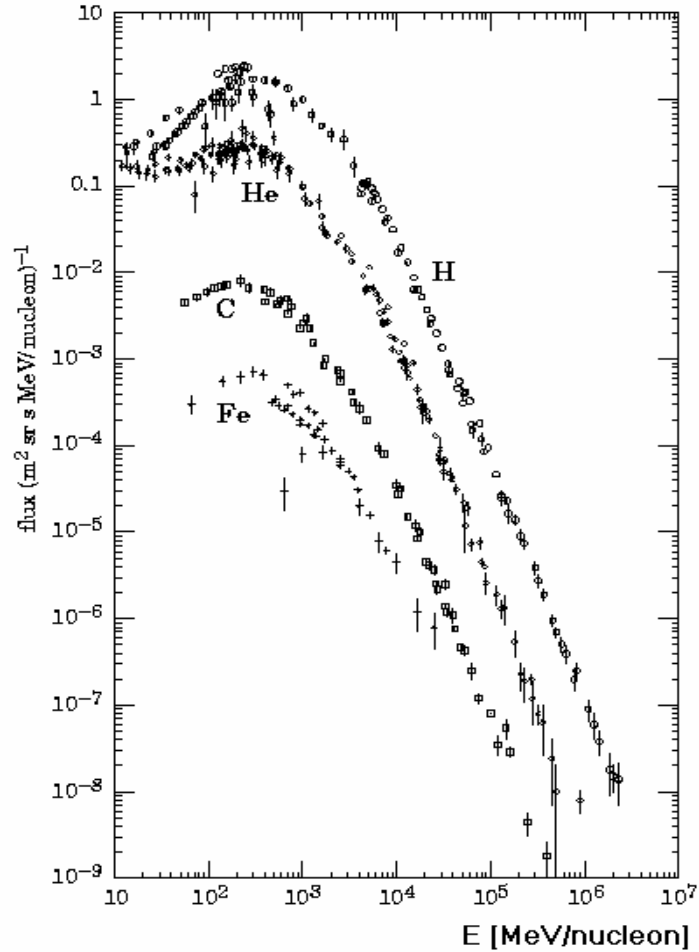

Fig 2 Major components of primary cosmic ray (indidual spectra)

The ground-based experiments detect extensive air showers (EAS), generated by interactions of the high-energy primary cosmic rays with the nuclei in the atmosphere. The cascade is initiated by primary cosmic ray particle when its first interaction with atmospheric nuclei occurs. A cascade of secondary particles is produced as a cosequence which degrade energy from the intial particle and depose their energy as well into the atmosphere. The developement of EAS into Earth amtosphere is preesented in fig.3. One can see the different components of the showers : the electromagnetic (soft) component, hadronic, muonic (hard) component and the Cherenkov light generated by charged particles in the shower.

A part of the secondary particles reach the ground at the observation level. The cosmic ray detection is made either by measuring of the energy passage as it is carried out by particles trough the atmosphere (the emissioin of Cherenkov light or the induced by nitrogen fluorescence light is a good example), or by direct detection in radiation detectors of particles which reach the ground. The arrival direction is usually assumed to the direction path in the atmosphere and has a resolution about of a degree. The indirect techniques are based on the measurements of one or few of the shower components and afterwards this is a matter of estimation the primary charge and energy. One of the main difficulties is that the obtained results are basically very model dependent. The reconstruction basically depends of the models of hadron-hadron interactions especially in the range of high energies. However, long duration experiments with large area can be operated on the ground and it is possible to registrate large number of events up to highest energies in the cosmic ray spectrum.

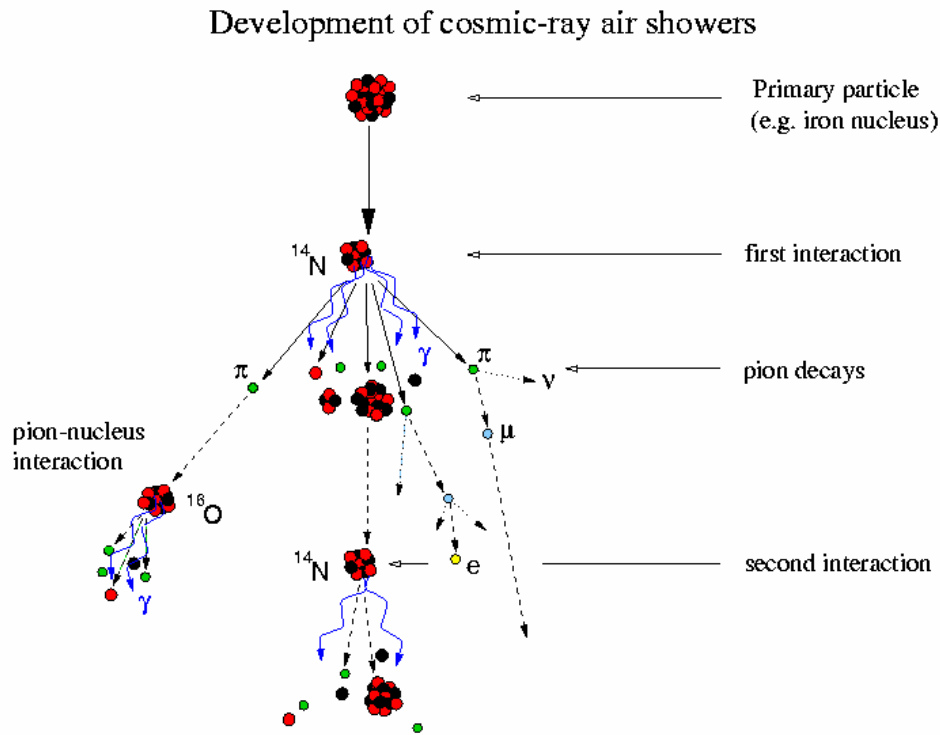

Fig. 3 Development of EAS induce by Iron primary nuclei

The particles of an extensive air shower (EAS) move nearly in the direction of the primary particles with velocities close to the velocity of the light. Transverse momenta of particles emitted in strong interactions and multiple scattering in the air produce a lateral dispersion fig.4. Differences in the velocities and, in particular, in the path lengths, when particles are traveling through the atmosphere, are the origin of a longitudinal dispersion (thickness) of the shower disk and of time delays of the arrival of the shower front, approximately represented by the relative arrival time of the first particle, relative to the arrival time of the core. The EAS thickness is manifested by the variation of the arrival times of the particles, observed at a particular fixed location of the EAS lateral extension while the variation of the delays with the distance from the shower centre reflects the shape (curvature) of the EAS front and the direction of the incidence.

In fig. 5 are presented the different techniques for registration of EAS. Usually the soft component (electron gamma) is measured at ground level using drift chambers or GM tubes. The scintillator detectors are used for registration of both soft and penetrating component. Deep underground detectors are used for registration of the penetrating component.

The registration of Cherenkov light is one of the most convenient techniques during the last years. One can reconstruct the image of the shower, method proposed by Weeks [9]. It is based of an array of photomultipliers at the focal plane of the telescope. The Cherenkov camera with good pixilation permits to discriminate the hadronic and electromagnetic showers and moreover to follow the development of the cascade into the atmosphere.

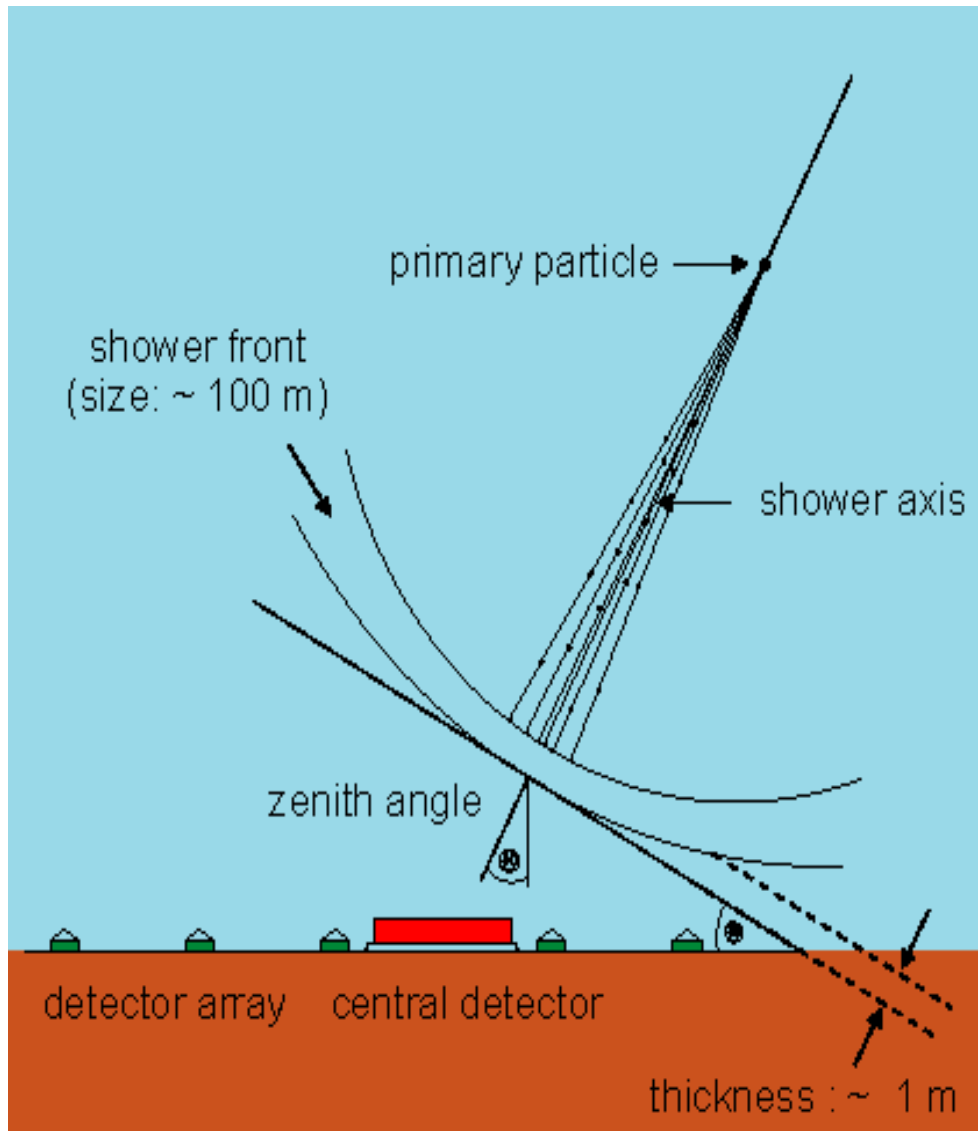

Fig.4 EAS within the shower front

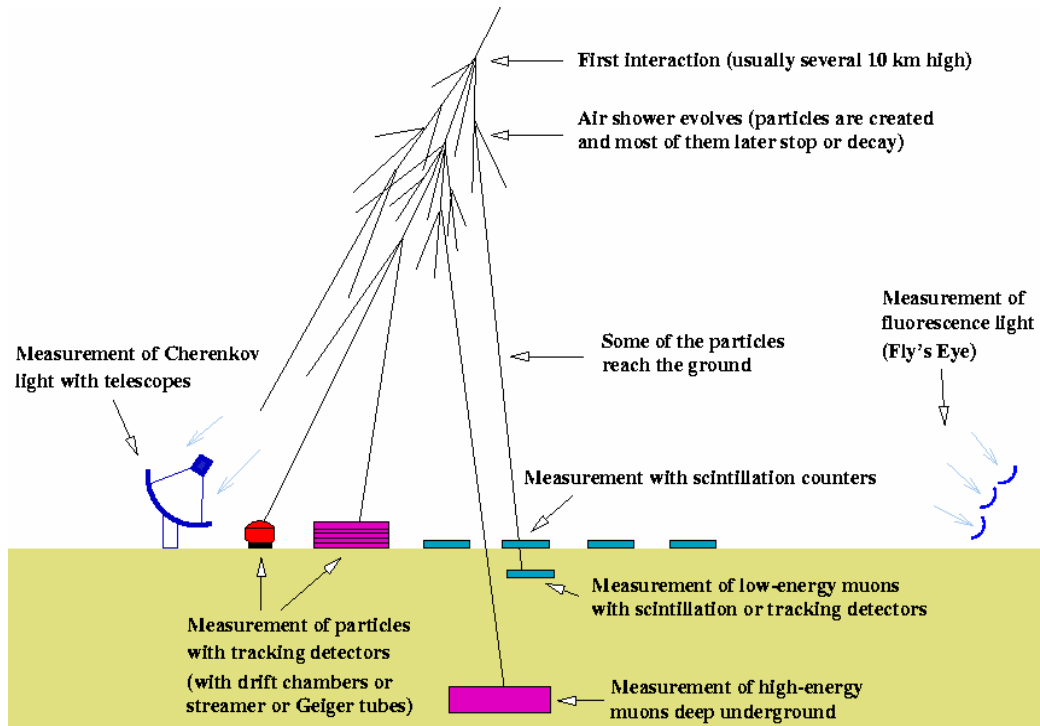

Fig.5 Different detection techniques for EAS registration

The Whiplle telescope used the technique of imaging the air shower to detect the Crab Nebula [10]. Thus this is a very power full tool for ground based gamma astronomy. The detection of the air Cherenkov light at ground level using an array of telescopes or photomultipliers contrary to the image technique is also a powerful tool for the both of the mentioned above problems - gamma astronomy [11, 12] and the all particle energy spectrum [13, 14]. In all the cases the atmospheric transparency is a crucial point for such type of investigations and the necessity for cloud less and moon less night. Thus the estimation of the atmospheric transparency and optical properties is very important. At the same time the influence of the atmospheric conditions on the development of the EAS is obvious. The influence of the near space is also very important.

Just as cosmic rays are deflected by the magnetic fields in interstellar space, they are also affected by the interplanetary magnetic field embedded in the solar wind (the plasma of ions and electrons blowing from the solar corona at about 400 km/sec), and therefore have difficulty reaching the inner solar system. Spacecraft venturing out towards the boundary of the solar system they have found that the intensity of galactic comic rays increases with distance from the Sun. As solar activity varies over the 11 year solar cycle the intensity of cosmic rays at Earth also varies, in anti-correlation with the sunspot number.

The Sun is also a sporadic source of cosmic ray nuclei and electrons that are accelerated by shock waves traveling through the corona, and by magnetic energy released in solar flares. During such occurrences the intensity of energetic particles in space can increase by a factor of $10^2$ to $10^6$ for hours to days. Such solar particle events are much more frequent during the active phase of the solar cycle. The maximum energy reached in solar particle events is typically 10 to 100 MeV, occasionally reaching 1 GeV (roughly once a year) to 10 GeV (roughly once a decade). Solar energetic particles can be used to measure the elemental and isotopic composition of the Sun, thereby complementing spectroscopic studies of solar material. At the same time such type of events play important role for the dose distribution especially on flight altitudes.

A third component of cosmic rays, comprised of only those elements that are difficult to ionize, including He, N, O, Ne, and Ar, was given the name "anomalous cosmic rays" because of its unusual composition. Anomalous cosmic rays originate from electrically-neutral interstellar particles that have entered the solar system unaffected by the magnetic field of the solar wind, been ionized, and then accelerated at the shock wave formed when the solar wind slows as a result of plowing into the interstellar gas, presently thought to occur somewhere between 75 and 100 AU from the Sun (one AU is the distance from the Sun to the Earth).

The connection between low energy cosmic ray and the Earth atmosphere is obvious, [15] it is a good basis for study of solar-terrestrial influences and space weather. Moreover the ability to forecast for long term space weather needs a precise knowledge of solar activity.
It is obvious that the space weather refers to conditions on the sun, solar wind and Earth's magnetosphere and ionosphere [16]. These conditions are connected and can impact the human activities.
Cosmic rays, especially galactic cosmic ray are essential part of the interstellar medium, their variations reflect on solar at all effects of the solar activity. Thus the relativistic cosmic ray, actually both solar and galactic registered and measured by neutron monitors can be used for space weather forecasting [17]. It is possible to study and forecast phenomena such as shocks and ground level enhancements and the magnetic properties of coronal mass ejection. Therefore the disturbances of the solar wind magnetosphere and the cosmic ray are related. They are generated by the same processes at the Sun. Several characteristic signatures in cosmic ray may be used for space weather applications [18] on the basis of neutron monitor data. Good examples are the solar proton events and Geomagnetic storms.
At the same time the variations of cosmic ray can be one of the possible links between the solar activity and the lower Earth atmosphere [17], precisely the large scale circulation of the lower atmosphere. They involve changes in the atmospheric transparency and cloud cover [18] due to the changes in stratospheric ionization produced by the secondary cosmic ray radiation. Thus the cosmic ray variation results on the high level cloud formation.
As was mentioned above the integral atmospheric transparency is related with cosmic ray variations and thus with the Forbes effect [21]. Moreover the transparency is one of the primary measures of the atmospheric state. It is obvious that long term series of atmospheric transparency measurements gives the possibility for quantitative estimate of the variability of air and therefore to make climatological conclusions with regard to contamination, cloud formation, humidity and radiative exchange.
With all this in mind we propose several activities at BEO Moussala and Alomar observatory towards to study the mentioned above effects.

## 2. Present status and future plans at BEO Moussala and Alomar observatory connected with correlation between cosmic ray variation and atmosphere parameters

Taking into account the unique characteristics of high mountain observatories they are privileged places and therefore they have been exploited during the years for cosmic ray and environmental studies. Generally the following specific objectives are pursued in attempt to provide basic information permitting analysis of the connection between cosmic ray variation and atmospheric parameters. The aim is the detailed, precise and contemporary measurements of cosmic ray intensity especially the muon, electron, gamma and neutron component and last but not least the atmospheric Cherenkov light. At the same time the atmosphere parameters, including anthropogenic products as well in different conditions of altitude, latitude and urban development is needed. Finally taking into account one of the principle objectives of BEOBAL project the collection of reliable data set obtained with intercalibrated instrumentation the final aim being to provide comparable results will be carried out.

Particularly the gamma and neutron energy spectra are evaluated in different laboratories using different methods. Appropriate techniques have been developed, due to the complexity of experimental neutron spectra and dose evaluation. Several Monte Carlo and semi-analytical codes are developed and implemented towards to study not only EAS (CORSIKA code is a good example [22]) but and the physics of the interactions between cosmic rays with aerosols and other atmospheric components. The experimental data collected in the laboratories can be used as input for the atmosphere composition and local ionization intensity. One of the final goals is the better and deeper understanding of the influence of galactic and solar cosmic rays, as well as the Sun activity in time, on the local and global climate variations as was mentioned previously.

Many atmospheric phenomena are related to cosmic ray ionization, that contribute to the electrical properties of the atmosphere: this effect is accepted in the mechanism of thundercloud and lighting production, in the ozone layer depletion, in sprites and elves ground to cloud triggering, in the charged particle precipitation. In addition, there are phenomenological evidences of the correlation between cosmic ray intensity variation and change in climate in the past long period. At present satellite observations show a correlation with the low cloud Earth cover: a possible microphysical mechanism could be attributed to the contribution of the ionization in the troposphere (3000-4000 m) in the cloud condensation nuclei formation.

Additionally the neutrons are produced by interaction of primary protons with atmosphere nuclei, and the neutron production rate and energy distribution strongly depend on the atmosphere physical characteristics, chemical composition, humidity, cloud density. In fact neutrons give an important contribution to the total dose in human exposure to cosmic radiation environment. Actually the obtained dose rate from secondary cosmic neutrons is the most important part of the dose produced by cosmic radiation. The final part of this proposal consists of the study of correlations between cosmic ray intensity and physical and chemical atmosphere parameters, to understand the influence of cosmic ray ionization on ozone variations and on microscopic aerosol and cloud condensation nuclei formation. Using new methods for data analysis and well-known methods and techniques is a good basis towards to the solution of the discussed problems.

*Measurements of secondary cosmic radiation*
The location of BEO Moussala is on the top of the highest mountain at Balkan peninsula. This is one of the most proper places in the region of Balkans and gives excellent possibility for high-mountain monitoring i.e. possibilities for measurements, experiments and monitoring for changes and processes at the atmosphere, pollution, air-transport, aerosol investigations, changes of gamma-background, dose-rate from neutron flux, cosmic ray investigations etc… At the same time the analyses of collected data gives information about relation between very different kind of parameters and factors.

A muon telescope is developed in INRNE and University of Blagoevgrad Bulgaria. The destination is the registration of secondary muons and measuring the cosmic rays variations. The original design of the telescope is based on 18 water Cherenkov detectors which are split in 2 slabs of 3x3 cells. The dimensions of the tanks are 50x50x12cm. The penetrating muons create Cherenkov photons registered by photomultipliers (fig.7). The telescope is under absorber so the electrons are rejected. In fact at Moussala we will use telescope with 8 water Cherenkov detectors (fig.6.).The distance between the two identical modules is 1m.

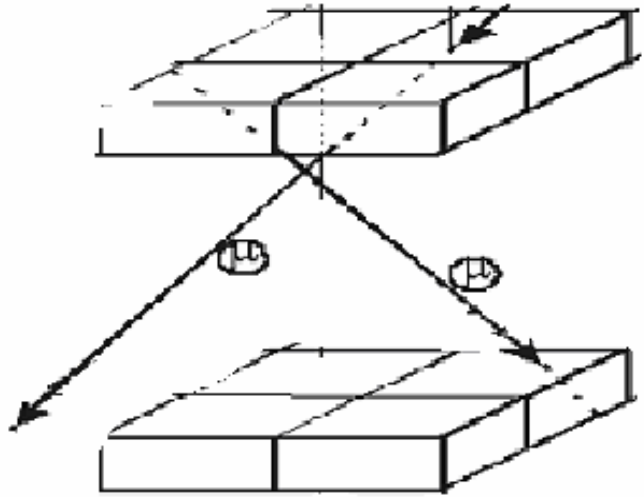

Fig.6 Muon Cherenkov telescope

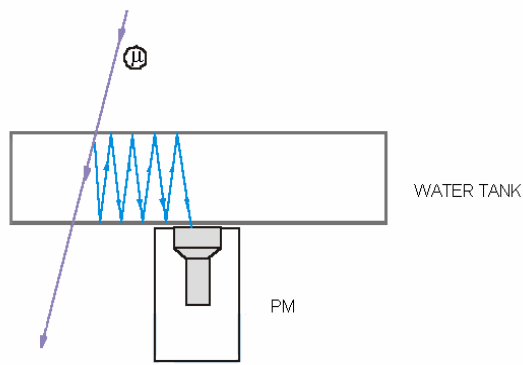

Fig.7 Water Cherenkov detector

One of the future projects yet in development is a neutron detector for absolute secondary cosmic neutron flux measurements. Such monitor will give good possibilities for analyzing the dose-rate of secondary neutron flux and average energy of integral spectra on secondary neutrons also after the reconstruction of the fluence of the secondary cosmic ray neutrons.

It will be based on Russian gas detectors filled with $BF_3$ of type SNM-15. The detectors will be situated under the roof of the building of BEO Moussala. On fig. 8 is shown the possible configuration and designs of the monitor for measurement of absolute neutron flux – group of 6 of detectors will be covered with glycerin moderator. This detector configuration is without lead i.e. is only with neutron moderator in this case glycerin. This is the main difference comparing to the usual neutron monitors. Thus the principle aim of this device is the measurement of the absolute neutron flux of secondary cosmic ray radiation. It is clear that the precise Monte Carlo modeling of the detector response is needed, the final aim being to estimate precisely the expected dose-rate. A preliminary experimental study of the detector efficiency using neutron flux is carried out. At present time the mechanical construction of the neutron flux-meter is done and the data acquisition system. Moreover a Monte Carlo with MCNP(x) code is carried out and the layer of the moderator is estimated.

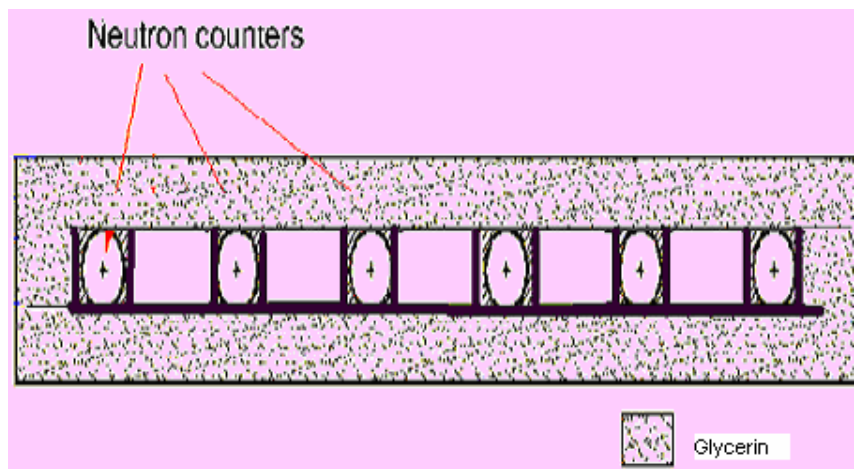

Fig. 8 Configuration of detectors of monitor for absolute neutron flux

With both of mentioned above devices it is possible to study the secondary cosmic ray flux and thus to estimate the variations of cosmic ray. In fact taking into account that the neutron flux is not sensitive to temperature variation s of the atmosphere both devices are complementary each other. Both, the muon telescope and the neutron flux-meter measure different components of the secondary cosmic ray radiation, which depends of different processes into atmosphere. The simultaneous measurements give excellent possibility to monitor the different processes.

At the same time at BEO Moussala one measure several atmospheric parameters such as temperature, humidity gas concentration of NOx and O3 and in a near future aerosol. Therefore the scientific potential seem to be huge and gives the possibility to study interesting phenomena such as the connection between cosmic ray variation and cloud cover and aerosol abundance in the atmosphere.

*Atmospheric transparency measurements*

As was mentioned above it exist possible connection between cosmic ray and atmospheric parameters. On the other hand the transparency is one of the primary measures of the atmospheric state and very important signature.

In this connection is very important to provide additional measurements of the integral atmospheric transparency. One of the possibilities is based on atmospheric Cherenkov light measurements using Cherenkov telescope. The telescope is made by two reflectors working in a coincidence scheme. The preliminary studies with CORSIKA 6.3 code [22] using GHEISHA [23] and QGSJET [24] hadronic interaction models and transparent (without any absorption) and atmosphere with Mie and Rayleigh scattering at Alomar observation level and Moussala observation level are presented in fig. 9. The used models are according [25]. Moreover such type of experiment gives the excellent possibility to study the different atmospheric profiles using additional measurement with LIDAR. Such investigation is carried out at Alomar observatory and the results will be published soon.

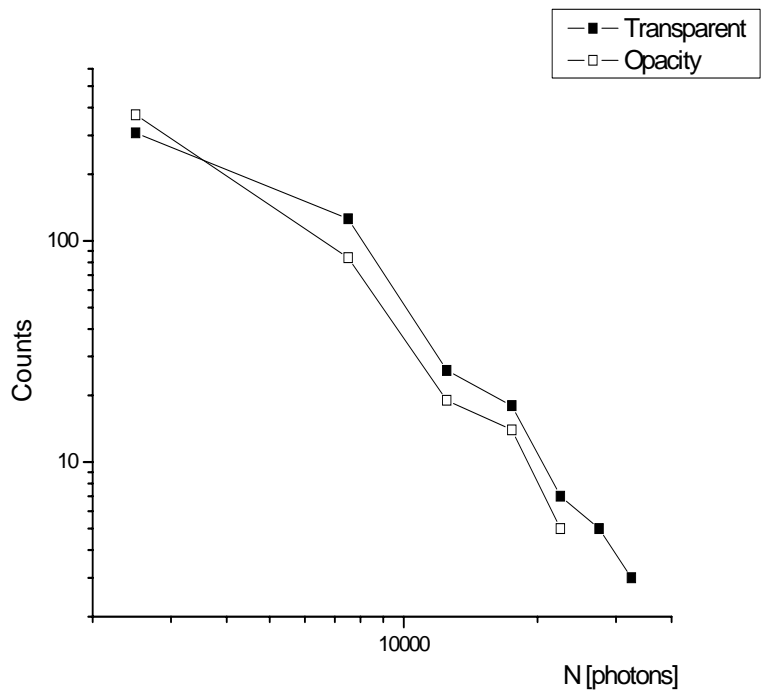

Fig.9 Amplitude spectra for different atmospheric profiles simulated with CORSIKA code

The third independent method for measuring the atmospheric transparency is the well known astrophotometric method using stellar standards and star light. Generally in optical astronomy an effect which must be corrected when calibrating instrumental magnitudes is the atmospheric extinction or the dimming of starlight by the terrestrial atmosphere. The longer the path length the starlight traverses through the atmosphere the more it is dimmed. Thus, a star close to the horizon will be dimmed more than one close to the zenith, and the observed brightness of a given star will change throughout a night, as its zenith distance varies. The path length through the atmosphere is known as the air mass. Consider an observation through the blanket of the atmosphere around the curved surface of the Earth. At any particular wavelength, $\lambda$ we can relate $m_0(\lambda)$, the magnitude of the observed object outside the atmosphere, to $m(\lambda)$, the magnitude of the observed object at the surface of the earth, by:

$$m(\lambda) = m_0(\lambda) + k(\lambda)X(z) \qquad (1)$$

where $X(z)$ is the air mass, $k(\lambda)$ is the extinction coefficient at wavelength $\lambda$ and $z$ is the zenith distance (the angular distance of the object from the zenith at the time of observation). $X$ is defined as the number of times the quantity of air seen along the line of sight is greater than the quantity of air in the direction of the zenith and will vary as the observed line of sight moves away from the zenith, that is, as $z$ increases. Note that the air mass is a normalized quantity and the air mass at the zenith is one.

For small zenith angles $X = \sec z$ is a reasonable approximation, but as $z$ increases, refraction effects, curvature of the atmosphere and variations of air density with height can become important. Hardie [26] gives a more refined relationship:

$$X = \sec z - 0.0018167(\sec z - 1) - 0.002875(\sec z - 1)^2 - 0.0008083(\sec z - 1)^3 \qquad (1)$$

and Young and Irvine [27] propose:

$$X = \sec z \left(1 - 0.0012(\sec^2 z - 1)\right). \qquad (3)$$

Both these equations imply the use of $z_t$, the true zenith angle, that is, the zenith angle to the observed object in the absence of the atmosphere as opposed to the apparent zenith angle $z_a$ affected by refraction effects.

The atmospheric extinction coefficient can be determined by observing the same object (through an appropriate filter) at several times during the night at varying zenith angles.

When the observed magnitudes of the object are plotted against computed air mass (fig.10), they should lie on a straight line with a slope equal to $k(\lambda)$. It is important to note that the extinction is dependent upon wavelength, being greater for blue light than red. Moreover the use of different filters (as example the standard UBV system with extended IR filters) can provide additional information about the atmosphere state. A relatively small telescope of about 20-25 cm mirror within the CCD camera will be enough for estimation of the extinction of BEO Moussala. The additional measurements with LIDAR can provide additional information.

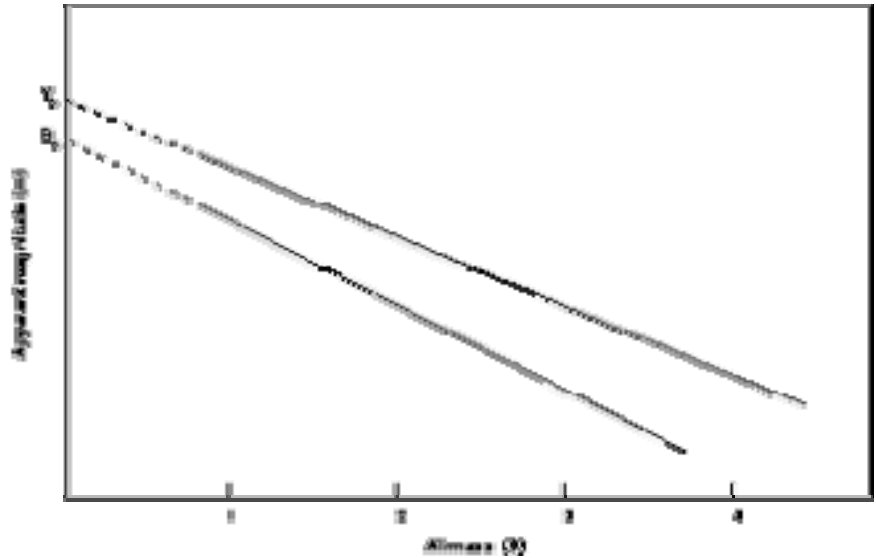

Fig. 10 Observed magnitudes of the object plotted against computed air mass

Generally two kinds of scattering are important: scattering by molecules of air, and scattering by solid particles or liquid droplets suspended in the air. Molecular scattering is usually called Rayleigh scattering, as it was first studied and explained by Lord Rayleigh. The suspended particles, on the other hand, are collectively known as aerosols, and their contribution is called aerosol scattering. Atmospheric aerosols are very diverse. They include tiny grains of mineral dust stirred up from the ground; particles of salt left when droplets of sea spray evaporate; bacteria, pollen grains, mold spores, and other "biosol" particles; photochemically produced droplets of sulfuric acid and other pollutants; soot particles produced in fires, and in vehicle exhaust; and many other materials. As most of these are produced at or near ground level, and are washed out of the atmosphere by condensation of cloud droplets on them, followed by precipitation, the aerosols all tend to be concentrated in the lowest part of the atmosphere; an exponential distribution with a scale height of about 1.5 km is a rough approximation to their vertical distribution. Because of their diversity, aerosol particles have a wide range of sizes. However, the ones most important for optical scattering turn out to be comparable to the wavelength being scattered, for typical size distributions. Therefore, as the wavelength decreases, we "see" the smaller particles better; the wavelength dependence of aerosol scattering is almost inversely proportional to the wavelength. (A $\lambda^{-1.2}$ power law is often used.) This produces considerable reddening. Most of the aerosol particles are so weakly absorbing that their extinction is almost entirely due to scattering, rather than absorption. However, soot (carbon) particles are quite strong absorbers, and a considerable part of their reddening is due to the increase in their absorption at short wavelengths. Absorption by molecules is sometimes called "true absorption," to emphasize its difference from extinction due to scattering; or "selective absorption," to emphasize its concentration in narrow spectral bands. The main absorbers in the visible spectrum are ozone (which absorbs in the Chappuis bands, in the orange part of the spectrum), water vapor (several bands in the longer-wavelength regions, noticed mainly under very humid conditions — hence the name "rain bands"), and oxygen (which produces Fraunhofer's A and B bands). Of these, the Chappuis bands of ozone are probably most important in green flashes, as they absorb strongly just in the wavelengths between red and green, and probably contribute to the abruptness of the color change seen in green flashes. The water-vapor bands are usually much less important. Thus it is easy to see the proposition at BEO Moussala to provide atmospheric transparency measurement with three different methods i.e. photometric standard using starlight, Cherenkov radiation and tropospheric LIDAR. At the same time within the aerosol measurements this will give excellent possibility to estimate the different contributions on the total atmospheric transparency. After that is a question of model predictions to estimate their impact on the environment.

*Muon hodoscope*

The muon hodoscope represents multi-channel device generally based on plastic scintillators. The original design of the hodoscope [28] is 512-channel large aperture muon hodoscope the aim being the investigation of solar-terrestrial physics. The estimated threshold of primary cosmic ray is 10 GeV. The estimated accuracy of measurement of cosmic ray muon directions is about 1-2 degree. The area of the hodoscope is 9m$^2$ and its counting rate is about a thousand events per second. The principle is shown in fig. 11.

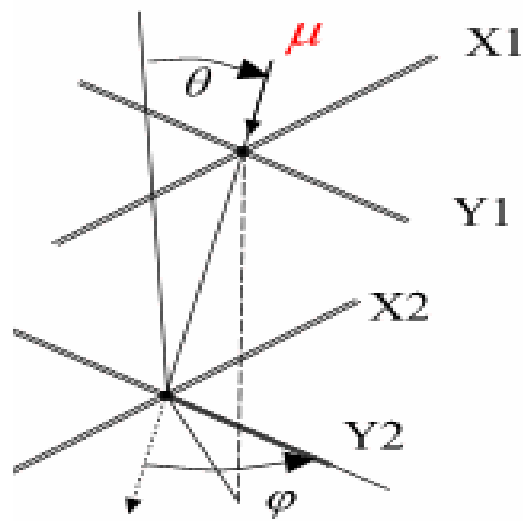

Fig. 11 Angles with detection principle of muon hodoscope

The detector includes 512 plastic scintillators counters. Each of the counters represents a narrow stripe; its length is 300 cm, cross-section is 2.5x1.0 cm$^2$. The photo multipliers coupled to each stripe provide the reliable detection of relativistic muons. The diameter of the PMT input window is 25mm. In the worst event, when a muon passes 3 meters away from PMT, it produces not less than 10 photoelectrons, which corresponds to 100% efficiency of the counter. The muon hodoscope is made of four layers. Each of the layers consists of 128 counters with a 2 mm iron sheet in front to reduce of the amount of knock-on electrons. The pict.1 shows the scheme of the muon hodoscope. The neighboring layers X, Y form the coordinate axes. The distance between the two pairs of layers is about 1 meter. To reject the soft component, the 5 cm thick lead filter is used. The fifth layer of scintillators (Z) is added for a master formation. The setup is located at the ground level and is capable of being oriented in the Sun's direction. The passage of a muon trough all layers causes the signals in four coordinate detectors. This provides the reconstruction of the angles with the accuracy of about 1-2 degrees. The aperture *delta omega* of the elementary cell is less than $5 \cdot 10^{-4}$ sr.

The scientific potential of the muon hodoscope is enormous. The Internal Gravitational Waves (IGW) registration where with the Muon hodoscope found on the barometric effect for muons. IGW are passing above the Muon hodoscope, and we are measuring of the intensity of muon streams for different angles in area = 30x30 km$^2$ in the moment, so we can observe the dimensional modulation of muon stream. This phenomenon had been registered first very authentically. IGW are transverse waves of density (unlike longitudinal acoustic waves). They arise in high the atmosphere (stratosphere). IGW can propagate along of the horizon and earthwards, but out of vertical large attenuation the above-ground barographs can not register it. We register IGW at the modulation of the muon stream of cosmic rays (The barometric effect for muons in the atmosphere).

Measurement of the temperature field along height of the atmosphere is founded on the temperature effect for muons (the zenithal angular muon spectrum depends on distribution of temperature in the atmosphere before height = 30 km). The temperature field is reconstructed of the solution of reverse problem with use of the muon spectrum measuring near the ground. The atmosphere is divided on many geopotential layers with fixed fluctuation of the temperature *delta $T_j(H_j)$*. Value of *delta $T_j$* defines from solution of the equations set.

The mean temperature (summed on all angles) see fig. 12; 2.different on the intensity of muons for varied angles. It is correspond to the temperature on varied geopotential levels of atmosphere (interval =50 - 200 bar). Quantity of layers restricts statistic precision of measurement of the intensity of muon stream (square of the device).

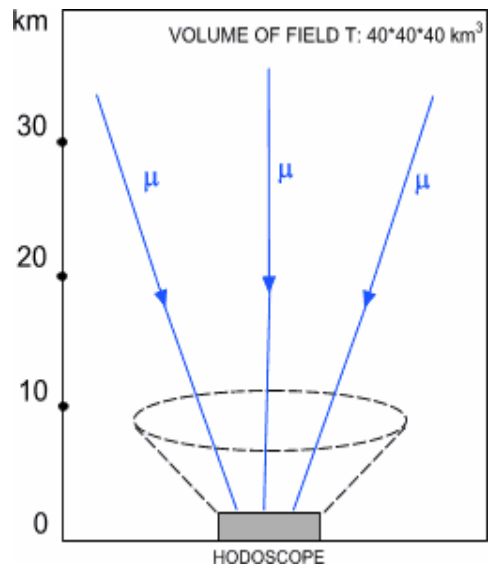

Fig. 12 Scanning of the atmospheric temperature with the muon hodoscope

*Monitoring of ozone layer*

We can obtain the distance estimate of variability of the ozone layer thickness on height 10-30 km from the value of difference of the effective night temperature of the atmosphere and the effective day temperature of the atmosphere on this height. The middle stratosphere heats owing to absorption of solar ultraviolet substantially. In the small thickness of ozone layer odds of the day and night temperature is small, but in the big thickness of ozone layer this odds is bigger. The preliminary model calculations agree with this idea. This problem can be resolved by the operations of frequent temperature measurements on stratosphere height during day over the time of all search time. Really, it will be make first wherewith the muon hodoscope in the operation of the continuous measurement of the angular distribution of cosmic muons. The hodoscope can make the continuous observation of the stratosphere ozone and control its season behavior and variability in dependence of a geophysical and technogenic factors. More detail information about this team will appear later.

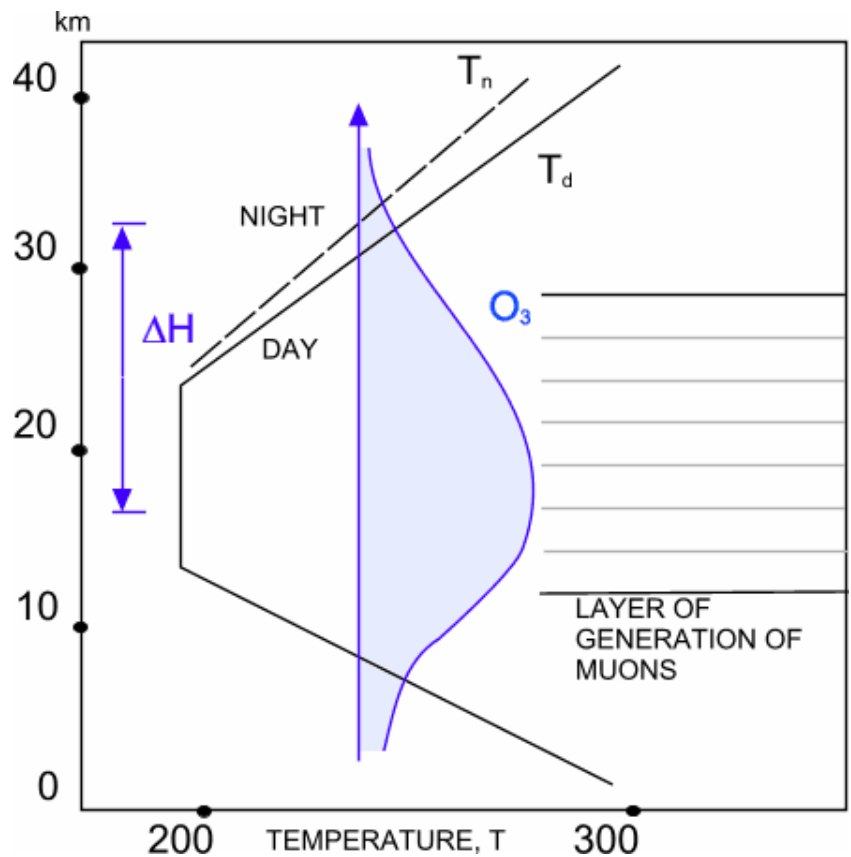

Fig. 13 Monitoring of the ozone layer

The further plans of BEO team are connected with similar design of the muon hodoscope. Are planning is to use water Cherenkov detectors, actually cylinder made by steal. The experience of our team for exploitation of water Cherenkov detectors is good. Moreover such type of the detector is cheaper. Several preliminary estimations using Monte Carlo technique are carried out. With the help of EGS4 code [29] actually modified version [30] the quantity of Cherenkov photons generated in the tank is estimated for several different water tanks. Moreover the registration efficiency estimation is obtained assuming simplified geometrical model. The attenuation of the Cherenkov light signal in the water is taken into account.

**Conclusion**

In this letter of intent are presented several of the present activities carried out at BEO Moussala and Alomar observatory connected with environmental studies, precisely the possible correlation between cosmic ray variation and atmosphere parameters. The scientific potential of several devices in development and future project is also discussed. Some preliminary estimations and expectations are presented.

**Acknowledgements**

We warmly acknowledge our colleagues from BEO Moussala and Alomar observatory. This work is supported under NATO grant EAP.RIG. 981843 and FP6 project BEOBAL.